\documentclass[%
 reprint,
 amsmath,amssymb,
 aps,
]{revtex4-2}
\usepackage{diagbox}
\usepackage{booktabs}
\usepackage{multirow}

\usepackage{graphicx}% Include figure files
\usepackage{dcolumn}% Align table columns on decimal point
\usepackage{bm}% bold math
\usepackage{hyperref}% add hypertext capabilities
\usepackage[mathlines]{lineno}% Enable numbering of text and display math
\usepackage{slashed}  
\usepackage{amsmath}
\usepackage{amssymb} 
\usepackage{natbib}
\usepackage{mathrsfs}
\usepackage{flushend}
\begin{document}

\preprint{APS/123-QED}

\title{Spatial sub-Rayleigh imaging via structured speckle illumination}% Force line breaks with \\

\author{Liming Li}%
 \email{liliming@sdut.edu.cn}
\affiliation{School of Physics and Optoelectronic Engineering, Shandong University of Technology, Zibo 255049, China\\}
\date{\today}% It is always \today, today,

\begin{abstract}
In a lens-assisted imaging scheme	with speckle illumination, the spatial resolution can surpass the Rayleigh resolution limit by a factor of $\sqrt{2}$ with second-order auto-correlation of light intensity. In this work, integrated with the nonlinear structured illumination after the speckle sinusoidally modulated, the second-order auto-correlation imaging can surpass the Rayleigh resolution limit by a factor of $2+\sqrt{2}$. In theory, a higher spatial resolution with the surpassing factor $N+\sqrt{N}$ is available by the $N$-order auto-correlation measurement. Our imaging scheme combined two super-resolution technologies not only enhances the spatial resolution of the lens-assisted imaging, but also promotes the practicality of the intensity correlation imaging.
\end{abstract}

\maketitle

%\tableofcontents

\section{Introduction}
Due to the wave property of light, the image of a point is a spot in a lens-assisted imaging system~\cite{BornWolf1999}. With the incoherent illumination, the correspondence of point-to-spot leads to a limited spatial resolution, known as the Rayleigh resolution limit $\Delta \rho_{\rm{R}}=0.61 \lambda/\text {NA}$~\cite{RayleighPM1879Investigations}, where $\lambda$ is the wavelength of light and NA is the numerical aperture of imaging system~\cite{Goodman2005}. In order to improve the spatial resolution with limited NA, some well-designed intelligent schemes have been invented in the field of optical microscopic imaging~\cite{HellOL1994Breaking,BetzigScience2006Imaging,RustNM2006Sub,HuangScience2008Three}. In 1999, Heintzmann and Cremer proposed that the high frequency information of text object less than twice the cut-off frequency lies inside the support of optical transfer function (OTF) by the aid of the illumination with a laterally modulated light\cite{HeintzmannSPIE1999Laterally}. In the following year  Gustafsson proposed the classical two-dimensional structured illumination microscopy (2D-SIM)\cite{GustafssonSPIE2000Doubling,GustafssonJMicrosc2000Surpassing} on the basis of Heintzmann's theory. After projecting sequentially three phase shifted sinusoidally fringes and collecting parallel raw images, people can separate those superposed frequency components by solving a group of linear equations\cite{GustafssonSPIE2000Doubling,LalIEEE2016Structured}. In theory, the resolution of 2D-SIM can surpass the Rayleigh resolution limit by a factor of 2. What's more, to acquire higher spatial resolution, Heintzmann proposed a non-linear SIM scheme\cite{HeintzmannJOSAA2002Saturated}, which was named the saturated structured illumination microscopy (SSIM). Laterly, some works demonstrated the super-resolution potential of SSIM\cite{GustafssonPNAS2005Nonlinear,RegoPNAS2012Nonlinear}.

From the multi-photon interference point of view, some sub-Rayleigh imaging schemes are developed via measurement the high-order correlation of light, such as the $N$-photon detection~\cite{GiovannettiPRA2009Sub,GuerrieriPRL2010Sub,XuAPL2015Experimental} and the optical centroid measurements~\cite{UnternährerOptica2018Super,TsangPRL2009Quantum,ToninelliOptica2019Resolution}. Due to weak mechanical structure and expensive detection equipment, the sub-Rayleigh imaging with quantum source is challenge to researchers in the experimental study. To avoid these obstacles, classical light were employed to acquire spatial high frequency information of text object via intensity correlated measurement~\cite{ZhangOE2015Sub,HongOE2017Heisenberg,LiOE2018Experimental,DertingerPNAS2009Fast,OhOL2013Sub,WangOL2015Spatial,DouOL2023Sub}. In fact, people need to find novel modulation schemes~\cite{ZhangOE2015Sub,HongOE2017Heisenberg,LiOE2018Experimental} or measure the auto-correlation of light intensity~\cite{DertingerPNAS2009Fast,OhOL2013Sub,WangOL2015Spatial,DouOL2023Sub}. For example, some imaing schemes using dynamic complex amplitude modulation can reach the standard quantum limit~\cite{ZhangOE2015Sub} and the Heisenberg limit~\cite{HongOE2017Heisenberg,LiOE2018Experimental}. Interestingly, there is a quite simple scheme to achieve sub-Rayleigh image. If the text object is illuminated with dynamic speckle, the resolution by the measurement with $N$-order auto-correlation of light can surpass the Rayleigh resolution limit by a factor of $\sqrt{N}$~\cite{DertingerPNAS2009Fast,OhOL2013Sub,WangOL2015Spatial,DouOL2023Sub,LiJO2019Beyond}.

In this work, we theoretically analyzed and numerical experimentally demonstrated a lens-assisted second-order sinusoidally structured speckle illumination microscopy (SSSIM) scheme, whose spatial resolution can surpass the Rayleigh resolution limit by a factor of $2+\sqrt{2}$. What's more, the surpassing factor of SSSIM scheme is $N+\sqrt{N}$ through $N$-order correlation of light.

\section{Theory}\label{Theory}

\begin{figure}[t]
	\centering
	\includegraphics[width=0.41\textwidth]{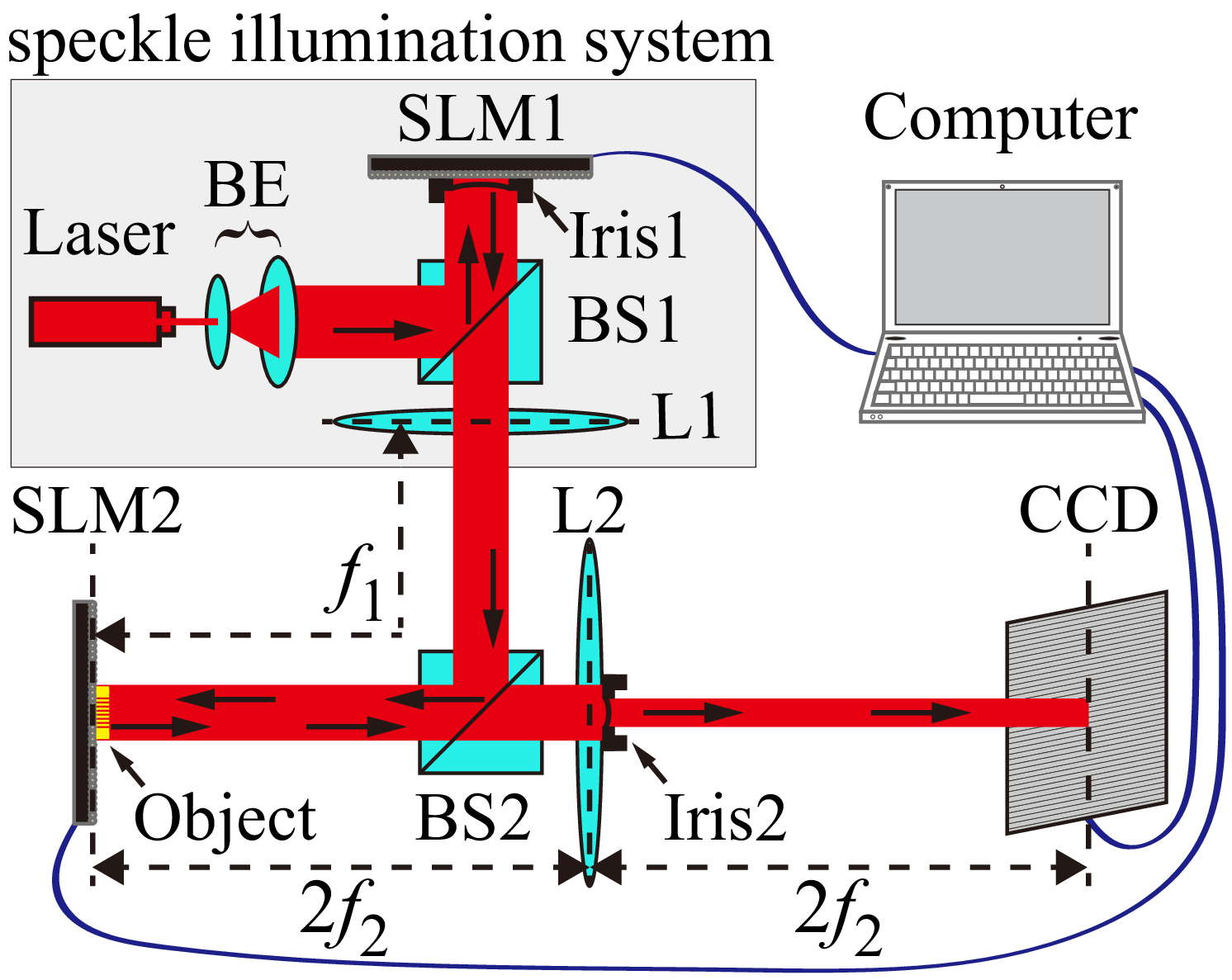}
	\caption{the experimental setup of SSSIM. BE: beam expander; BS1, BS2: beam splitter; SLM1, SLM2: spatial light modulators; L1, L2: lenses; CCD: charge-coupled device.} \label{scheme}
\end{figure}

Figure~\ref{scheme} depicts the SSSIM experimental setup, which includes a $2f-2f$ imaging system and a controllable dynamic speckle illumination system. A continuous-wave $\lambda$ = 632.8 nm single-mode laser beam was expanded and collimated by a beam expander (BE), and was then reflected by a 50:50 non-polarized beam splitter (BS1) onto a phase-only reflective spatial light modulator (SLM1, an element pixel size of 24.6 × 24.6 $\mu \text{m}^{2}$ and a total pixels 512 $\times$ 512). The reflected beam from the SLM1 transmitted through the BS1 again and was then refracted by a lens (L1) with a focal length $f_1$ = 80 cm. Here, a amplitude-only reflective SLM2 (an element pixel size of 20 × 20 $\mu \text{m}^{2}$ and a total pixels 512 $\times$ 512) was put on the rear focal plane of L1. Therefore, the dynamic speckle illumination can be realize on the plane of SLM2 while the random phase pattern was encoded onto the wavefront of light by reflected from the SLM1~\cite{LiPRA2019Transverse,LiPLA2021Eigenmode,WuCPB2024High}. In addition, it's easy to control the size of speckle particle by adjusting the diameter $\Delta L_1$ of iris1, which was placed as close as possible to the SLM1. After passing through L1, the beam was reflected by another 50:50 non-polarized beam splitter (BS2) and was incident normally on the SLM2. The reflected beam from the SLM2 transmitted through the BS2 again and another lens (L2), and was then recorded by using a charge coupled device (CCD) camera. The SLM2, the L2 and the CCD camera create a $2f_2-2f_2$ imaging system, where $f_2$ = 30 cm is the focal length of L2. The text object put right behind the SLM2. Here, another iris2 was placed as close as possible to L2, and its diameter $\Delta L_2$ = 0.46 mm decides the Rayleigh resolution limit of the imaging system. As we can see, a dynamically sinusoidally structured speckle illumination system can be realized with help of the SLM2.

When the size of speckle particle on the plane of text object is small than the Rayleigh resolution limit of the $2f_2-2f_2$ imaging system, the first-order intensity distribution on the image plane can be expressed as~\cite{WangOL2015Spatial}:
\begin{equation}\label{EQ1}
I({{\vec \rho _{\rm{i}}}}) \propto \int\limits_T {d{\vec \rho _{\rm{o}}}{T^2}({{\vec \rho _{\rm{o}}}})}H( {{\vec \rho _{\rm{i}}} - {\vec \rho _{\rm{o}}}}),
\end{equation}
where the point-spread function (PSF) of imaging system
\begin{equation}\label{EQ2}
	H( {{\vec \rho _{\rm{i}}} - {\vec \rho _{\rm{o}}}} )={\rm{som}}{{\rm{b}}^2} ( {k{\rm{NA}}\left| {{\vec \rho _{\rm{i}}} - {\vec \rho _{\rm{o}}}} \right|} ).
\end{equation}
Here, $\vec \rho _{\rm{i}}$ and $\vec \rho _{\rm{o}}$ are the transverse coordinates on the image plane and object plane, respectively. $T(\vec \rho _{\rm{o}} )$ is the aperture function of the text object. In addition, $\text{somb}(\rho)=J_1(\rho)/\rho$ is the Airy function in Eq.~(\ref{EQ2}), $\text{NA} = \Delta L_2 /(4f_2)$, and $k=2\pi/\lambda$ is the wave number.

\begin{figure}[t]
	\centering
	\includegraphics[width=0.40\textwidth]{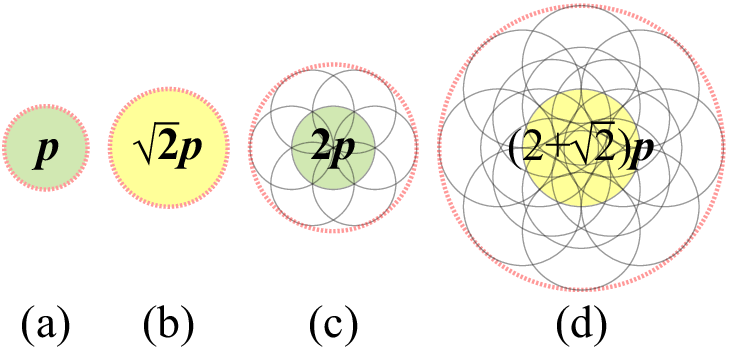}
	\caption{OTF support. (a) the first-order intensity imaging scheme; (b) the second-order auto-correlation imaging scheme; (c) the classical 2D-SIM; (d) the SSSIM scheme.} \label{OTF}
\end{figure}

What's more, the second-order fluctuation auto-correlation function on the image plane is given by~\cite{WangOL2015Spatial}:
\begin{equation}\label{EQ3}
	\Delta{G^{(2)}}( {{\vec \rho_{\rm{i}}}})\propto \int\limits_T {d{\vec \rho _{\rm{o}}}{T^4}({{\vec \rho _{\rm{o}}}})} H^2( {{\vec \rho _{\rm{i}}} - {\vec \rho _{\rm{o}}}}).
\end{equation}
Due to narrower the full width at half maximum (FWHM) of $H^2( {{\vec \rho _{\rm{i}}} - {\vec \rho _{\rm{o}}}})$ than $H({{\vec \rho _{\rm{i}}} - {\vec \rho _{\rm{o}}}})$ the PSF of the first-order intensity imaging system, the circle of OTF support with the second-order auto-correlation of light is $\sqrt2p$, where $p=1/\Delta \rho_{\rm{R}}$ is the circle of OTF support of the first-order intensity imaging. Figure~\ref{OTF} shows four kinds of OTF supports of imaging scheme with identical NA. The OTF support of the first-order intensity imaging is shown in Fig.~\ref{OTF}(a). Comparing with $\sqrt2p$ the case of second-order auto-correlation imaging as shown in Fig.~\ref{OTF}(b), the circle of OTF support with classical 2D-SIM is $2p$ as shown in Fig.~\ref{OTF}(c)~\cite{LalIEEE2016Structured}. So, what would happen on the circle of OTF support in the SSSIM scheme?

To realize the frequency expansion in the SSSIM scheme, the illumination speckle are modulated by SLM2 with a series of sinusoidal intensity patterns 
\begin{equation}\label{EQ4}
	S{I_{\varphi ,\phi }}( {{\vec \rho _{\rm{o}}}} )=1-\cos  (2\pi{{\vec p}_\varphi } \cdot {{\vec \rho }_{\rm{o}}}+\phi  ),
\end{equation}
where $\varphi \in \text{A}=[0, \pi/4, 2\pi/4, 3\pi/4]$ indicates the orientation and $\phi\in \text{B}=[0, 2\pi/5,4\pi/5,6\pi/5,8\pi/5]$ is the phase of sinusoidal pattern, ${\vec \rho }_{\rm{o}} = ( {{x_\text{o}},{y_\text{o}}} )$ is the coordinates of object plane, ${\vec p}_\varphi=( p\cos\varphi, p\sin\varphi)$ is sinusoidal illumination frequency vector in reciprocal space, respectively. The raw-images of the second-order fluctuation auto-correlation function on the image plane with speckle illumination sinusoidally modulated can be expressed as:
\begin{equation}\label{EQ5}
\Delta G_{\varphi ,\phi }^{(2)}({{\vec \rho_{\rm{i}}}})=[ {SI_{\varphi ,\phi }^2({{\vec \rho _{\rm{o}}}}){T^4}({{\vec \rho _{\rm{o}}}})}]\otimes H^2({{\vec \rho_{\rm{i}}}- {\vec \rho_{\rm{o}}}}),
\end{equation}
where $\otimes$ is convolution operator. By using convolution theorem~\cite{Goodman1995Introduction} for Eq.~(\ref{EQ5}), the Fourier transform of observed image is given by:
\begin{equation}\label{EQ6}
\begin{aligned}
\begin{array}{l}
	\widetilde{\Delta G_{\varphi ,\phi }^{(2)}}(k) =[{\widetilde{SI_{\varphi ,\phi }^2}(k) \otimes\widetilde{T^4}(k)}]{\widetilde{H^2}}(k)\\[4pt]
	= \left[ \begin{array}{l}
		{\frac{3}{2}}\widetilde{T^4}(k)\\[2pt]
		-{e^{i\phi }}\widetilde{T^4}({k-{{\vec p}_\varphi } \cdot{{\vec\rho }_{\rm{o}}}})\\[2pt]
		-{e^{-i\phi }}\widetilde{T^4}({k+{{\vec p}_\varphi } \cdot {{\vec\rho }_{\rm{o}}}})\\[2pt]
		+\frac{1}{4}{e^{2i\phi }}\widetilde{T^4}({k-2{{\vec p}_\varphi}\cdot{{\vec\rho }_{\rm{o}}}})\\[2pt]
		+\frac{1}{4}{e^{-2i\phi}}\widetilde{T^4}({k+2{{\vec p}_\varphi}\cdot {{\vec\rho}_{\rm{o}}}})
	\end{array} \right]{\widetilde{H^2}}( k ),
\end{array}
\end{aligned}
\end{equation}
where
\begin{equation}\label{EQ7}
	\begin{array}{l}
		\widetilde{SI_{\varphi ,\phi }^2}(k) =\frac{3}{2}\delta(k)\\[4pt]
		-{e^{i\phi}}\delta({k-{{\vec p}_\varphi}\cdot {{\vec\rho }_{\rm{o}}}})
		-{e^{-i\phi }}\delta({k+{{\vec p}_\varphi} \cdot {{\vec \rho }_{\rm{o}}}} )\\[4pt]
		+\frac{1}{4}{e^{2i\phi}}\delta ({k - 2{{\vec p}_\varphi } \cdot {{\vec \rho}_{\rm{o}}}} ) 
		+\frac{1}{4}{e^{-2i\phi }}\delta ({k+2{{\vec p}_\varphi }\cdot{{\vec \rho}_{\rm{o}}}} )
	\end{array}
\end{equation}
decides the frequency shift. Comparing with $\widetilde{T^4}(k)$ the zero-fold frequency shift domain ($\text{FSD}_0$) in Eq.~(\ref{EQ6}), $\widetilde{T^4}({k-{{\vec p}_\varphi } \cdot {{\vec \rho }_{\rm{o}}}})$, $\widetilde{T^4}( {k + {{\vec p}_\varphi } \cdot {{\vec \rho }_{\rm{o}}}})$, $\widetilde{T^4}( {k - 2{{\vec p}_\varphi } \cdot {{\vec \rho }_{\rm{o}}}})$ and $\widetilde{T^4}( {k + 2{{\vec p}_\varphi } \cdot {{\vec \rho }_{\rm{o}}}})$ are $\text{FSD}_1$, $\text{FSD}_{-1}$, $\text{FSD}_2$ and $\text{FSD}_{-2}$, respectively. Although superposed together, those five FSDs can be separated by the linear equation:
\begin{widetext}
\begin{equation}\label{EQ8}
\left[ \begin{array}{l}
\widetilde{T^4}  ( k ){{\tilde H}^2}( k )\\[5pt]
\widetilde{T^4}  ( {k - {{\vec p}_\varphi } \cdot {{\vec \rho }_{\rm{o}}}} ){{\tilde H}^2}( k )\\[5pt]
\widetilde{T^4}  ( {k + {{\vec p}_\varphi } \cdot {{\vec \rho }_{\rm{o}}}} ){{\tilde H}^2}( k )\\[5pt]
\widetilde{T^4}  ( {k - 2{{\vec p}_\varphi } \cdot {{\vec \rho }_{\rm{o}}}} ){{\tilde H}^2}( k )\\[5pt]
\widetilde{T^4}  ( {k + 2{{\vec p}_\varphi } \cdot {{\vec \rho }_{\rm{o}}}} ){{\tilde H}^2}( k )
\end{array} \right] = {\left[ \begin{array}{l}
		\frac{3}{2}\quad  - {e^{i{\phi _1}}}\quad  - {e^{ - i{\phi _1}}}\quad \frac{1}{4}{e^{2i{\phi _1}}}\quad \frac{1}{4}{e^{ - 2i{\phi _1}}}\\[5pt]
		\frac{3}{2}\quad  - {e^{i{\phi _2}}}\quad  - {e^{ - i{\phi _2}}}\quad \frac{1}{4}{e^{2i{\phi _2}}}\quad \frac{1}{4}{e^{ - 2i{\phi _2}}}\\[5pt]
		\frac{3}{2}\quad  - {e^{i{\phi _3}}}\quad  - {e^{ - i{\phi _3}}}\quad \frac{1}{4}{e^{2i{\phi _3}}}\quad \frac{1}{4}{e^{ - 2i{\phi _3}}}\\[5pt]
		\frac{3}{2}\quad  - {e^{i{\phi _4}}}\quad  - {e^{ - i{\phi _4}}}\quad \frac{1}{4}{e^{2i{\phi _4}}}\quad \frac{1}{4}{e^{ - 2i{\phi _4}}}\\[5pt]
		\frac{3}{2}\quad  - {e^{i{\phi _5}}}\quad  - {e^{ - i{\phi _5}}}\quad \frac{1}{4}{e^{2i{\phi _5}}}\quad \frac{1}{4}{e^{ - 2i{\phi _5}}}
	\end{array} \right]^{ - 1}}\left[ \begin{array}{l}
	\widetilde{\Delta G_{\varphi ,\phi_1 }^{( 2 )}} (k)\\[2pt]
	\widetilde{\Delta G_{\varphi ,\phi_2 }^{( 2 )}} (k)\\[2pt]
	\widetilde{\Delta G_{\varphi ,\phi_3 }^{( 2 )}} (k)\\[2pt]
	\widetilde{\Delta G_{\varphi ,\phi_4 }^{( 2 )}} (k)\\[2pt]
	\widetilde{\Delta G_{\varphi ,\phi_5 }^{( 2 )}} (k)\\
\end{array} \right],
\end{equation}
\end{widetext}
where $[M]^{-1}$ means the inverse matrix of $M$. As we can see, five independent measurements with different $\phi$ can give those five $\text{FSD}$s in the orientation of $\varphi$. With the frequency expansion at four orientations, the circle of OTF support of second-order SSSIM is $(2+\sqrt2)p$ as shown in Fig.~\ref{OTF}(d). Considering the similarity of frequency operations (Shift, Merge and Inverse Fourier transformation) with 2D-SIM, we will not elaborate on it here. Please refer to the 2D-SIM~\cite{LalIEEE2016Structured}. In the following section, we will verify the spatial high-resolution of the SSSIM scheme.
\section{Numerical Experiment} \label{EandNS}
In order to test the spatial resolution of SSSIM, we perform the follow-up numerical experiment. To create the dynamic speckle illumination system, the phase structure $e^{i\theta (\rho_\text{s},t)}$ were loaded on the SLM1, where the phasor $\theta (\rho_\text{s},t)\in[0, 2\pi)$ is completely random in the spatial $\rho_\text{s}$ and the time $t$ distribution. The average size $l_c$ of speckle particle (i.e. the transverse coherence length) on the object plane is $ \lambda f_1 / \Delta L_1=$ 85 $\mu \text{m}$, where the diameter of iris1 $\Delta L_1$ is 6 mm. What's more, the Rayleigh-resolution limit $\Delta \rho_{\rm{R}}$ of the $2f_2-2f_2$ imaging system is 1 $\text{mm}$. Therefore, the average size of speckle particle meets the requirement of auto-correlation imaging scheme.

\begin{figure}[t]
	\centering
	\includegraphics[width=0.49\textwidth]{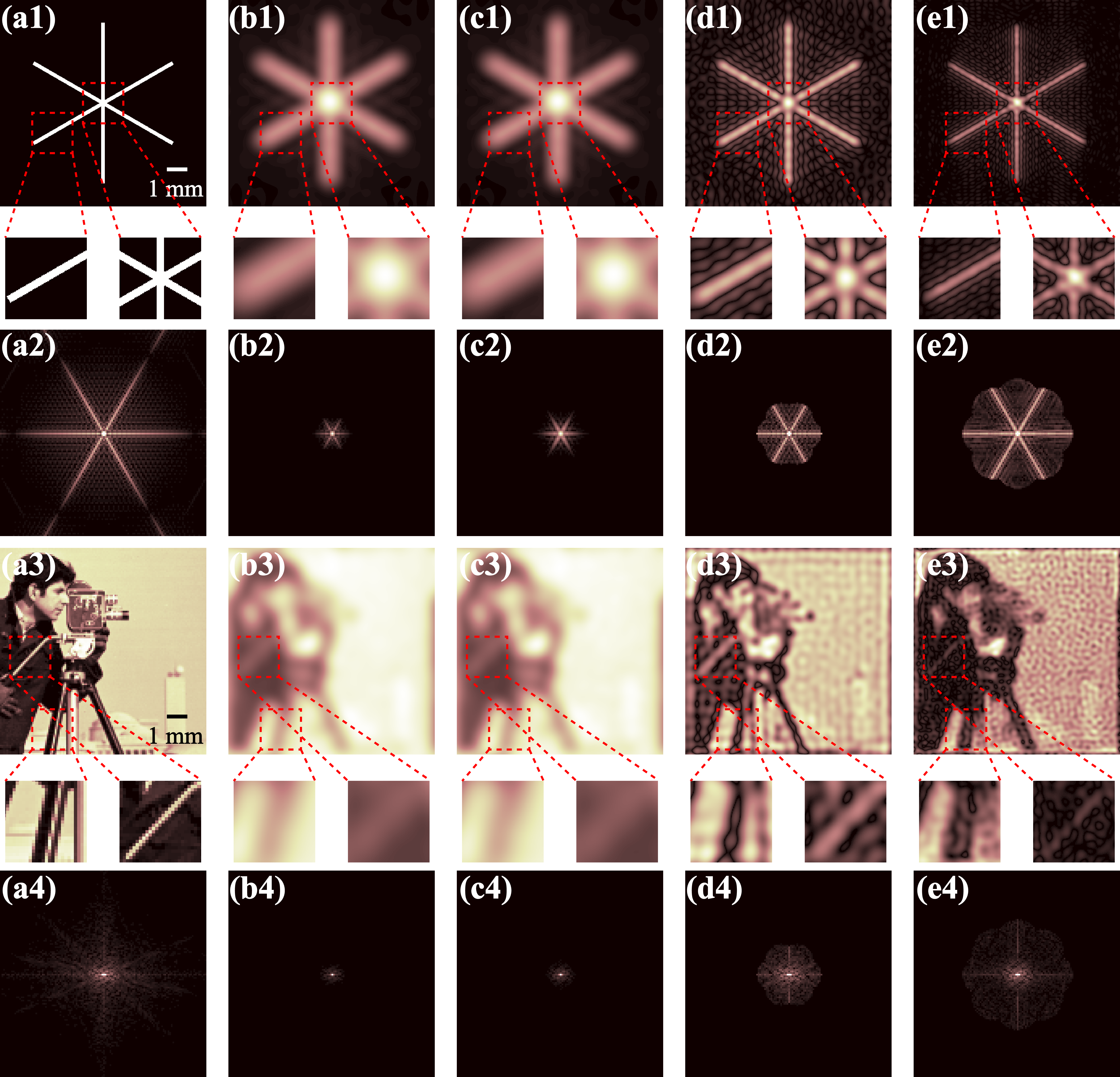}
	\caption{Test objects: (a1), (a3); the first-order intensity imaging: (b1), (b3); the second-order auto-correlation imaging: (c1), (c3); 2D-SIM: (d1), (d3); SSSIM: (e1), (e3); (a2-e2, a4-e4) Fourier domain of images in (a1-e1, a3-e3), respectively. Those insets show the enlarged details of images.}\label{SSSIM}
\end{figure}

Figure~\ref{SSSIM} gives the numerical experiment with a 0-1 binary text object '$\ast $' and a natural grayscale text object 'Cameraman'. The number of total frame is 10000 for the second-order auto-correlation imaging and those raw images in the SSSIM scheme. Figures~\ref{SSSIM} (a1, a3) show those test objects, which are 512×512 pixels. The imaging results of first-order intensity imaging and second-order auto-correlation imaging are shown in Figs.~\ref{SSSIM}(b1, b3) and Figs.~\ref{SSSIM}(c1, c3), respectively. Although the circle of OTF support of auto-correlation imaging scheme is larger than that of the first-order intensity imaging, the improvement of spatial resolution is inapparent. However, as the frequency detection domain further expanded, those imaging results will present more details. The imaging results of 2D-SIM and SSSIM are shown in Figs.~\ref{SSSIM}(d1, d3) and Figs.~\ref{SSSIM}(e1, e3), respectively. Herein the 2D-SIM scheme, the orientation and the phase of sinusoidal illumination pattern satisfy $\varphi \in [0, \pi/3, 2\pi/3]$ and $\phi\in [0, 2\pi/3,4\pi/3]$, repectively. As we can see, those imaging results with 2D-SIM scheme and SSSIM scheme show significant improvement on the spatial resolution whether 0-1 binary object or natural grayscale object. From those details of images shown in the insets, we can further find that the resolution of SSSIM is higher than the 2D-SIM. Figures~\ref{SSSIM}(a2-e2, a4-e4) present the Fourier domain of images in Figs.\ref{SSSIM} (a1-e1, a3-e3), respectively. As we can see, those Fourier domains in Figs.~\ref{SSSIM}(b2-e2, b4-e4) are similar with the corresponding OTF support as shown in Fig.~\ref{OTF}.

The fidelity of imaging result can be estimated by calculating the peak signal-to-noise ratio (PSNR):
\begin{equation}\label{EQ9}
	\begin{split}
		\text{PSNR} = 10\log_{10}{[ {{( {2^{\text{BIT}} - 1} )}^2}/\text{MSE} ]},
	\end{split}
\end{equation}
where $\text{BIT}=8$ is the bit depth of images and $\text{MSE}$ is the mean-square error of the imaging result with respect to the test object. Here, $\text{MSE}$ is defined as:
\begin{equation}\label{EQ10}
	\begin{split}
		\text{MSE}=\frac{1}{\tau^2} \sum\limits_{n_1=1}^\tau\sum\limits_{n_2=1}^\tau {\left[I(n_1,n_2)-T(n_1,n_2) \right]^2},
	\end{split}
\end{equation}
where $\tau=512$ is the total number of pixels in single dimension. The imaging result $I(n_1,n_2)$ and the test object $T(n_1,n_2)$ are digitized to $0\sim255$. It can be seen that the higher the quality of the result, the greater the PSNR. According to numerical results, PSNR of Figs.~\ref{SSSIM}(b1-e1) with the 0-1 binary object are 14.74 dB, 15.20 dB, 17.82 dB and 16.50 dB, respectively. What's more, PSNR of Figs.~\ref{SSSIM}(b3-e3) with the garyscale object are 5.47 dB, 5.52 dB, 16.78 dB and 16.39 dB, respectively. It can be seen that the quality of image results has been optimized through the 2D-SIM scheme and the SSSIM scheme. Although the SSSIM scheme acquires more frequency information and displays more object details, the image quality is slightly lower than the 2D-SIM scheme. Considering that those imaging results with the SSSIM scheme originated from the frequency merge by 25 FSDs and the number is 9 for 2D-SIM, the frequency merge leads to substantial ringingeffect so that a serious loss in imaging quality. If denoising and filtering techniques of 2D-SIM~\cite{Gustafsson2008Three-dimensional,Müller2016Opensource,Wen2021High,Smith2021Structured} can be appropriately introduced, the imaging quality will be improved and the imaging results can show more details of text object.

\section{Discussion} \label{Discussion}
In theory, the spatial resolution of SSSIM can be further improved through high-order correlation of light. The $N$-order fluctuation auto-correlation function on the image plane with speckle illumination sinusoidally modulated can be expressed as:
\begin{equation}\label{EQ11}
\Delta G_{\varphi ,\phi }^{(N)}({{\vec\rho_{\rm{i}}}})=[ {SI_{\varphi ,\phi }^N({{\vec\rho _{\rm{o}}}}){T^{2N}}({{\vec\rho _{\rm{o}}}})}]\otimes H^N({{\vec\rho_{\rm{i}}}- {\vec\rho_{\rm{o}}}}),
\end{equation}
where the modulation term ${SI_{\varphi ,\phi }^{N}({{\vec\rho _{\rm{o}}}})}$ desides FSD of SSSIM. Here, those Fourier transform of modulation terms with $N=3$ and $N=4$ are expressed as:
\begin{equation}\label{EQ12}
	\begin{array}{l}
		\widetilde{SI_{\varphi ,\phi }^3}(k) =\frac{5}{2}\delta(k)\\[4pt]
		-\frac{15}{8}{e^{i\phi}}\delta({k-{{\vec p}_\varphi}\cdot {{\vec\rho }_{\rm{o}}}})
		-\frac{15}{8}{e^{-i\phi}}\delta({k+{{\vec p}_\varphi}\cdot {{\vec\rho }_{\rm{o}}}})\\[4pt]
				
		+\frac{3}{4}{e^{2i\phi}}\delta ({k - 2{{\vec p}_\varphi } \cdot {{\vec \rho}_{\rm{o}}}} ) 
		+\frac{3}{4}{e^{-2i\phi}}\delta ({k + 2{{\vec p}_\varphi } \cdot {{\vec \rho}_{\rm{o}}}} )\\[4pt]
		
		-\frac{1}{8}{e^{3i\phi}}\delta ({k - 3{{\vec p}_\varphi } \cdot {{\vec \rho}_{\rm{o}}}} ) 
		-\frac{1}{8}{e^{-3i\phi}}\delta ({k + 3{{\vec p}_\varphi } \cdot {{\vec \rho}_{\rm{o}}}} )\\
	\end{array}
\end{equation}
and
\begin{equation}\label{EQ13}
	\begin{array}{l}
		\widetilde{SI_{\varphi ,\phi }^4}(k) =\frac{35}{8}\delta(k)\\[4pt]
		-\frac{7}{2}{e^{i\phi}}\delta({k-{{\vec p}_\varphi}\cdot {{\vec\rho }_{\rm{o}}}})
		-\frac{7}{2}{e^{-i\phi}}\delta({k+{{\vec p}_\varphi}\cdot {{\vec\rho }_{\rm{o}}}})\\[4pt]
		
		+\frac{7}{4}{e^{2i\phi}}\delta ({k - 2{{\vec p}_\varphi } \cdot {{\vec \rho}_{\rm{o}}}} ) 
		+\frac{7}{4}{e^{-2i\phi}}\delta ({k + 2{{\vec p}_\varphi } \cdot {{\vec \rho}_{\rm{o}}}} )\\[4pt]
		
		-\frac{1}{2}{e^{3i\phi}}\delta ({k - 3{{\vec p}_\varphi } \cdot {{\vec \rho}_{\rm{o}}}} ) 
		-\frac{1}{2}{e^{-3i\phi}}\delta ({k + 3{{\vec p}_\varphi } \cdot {{\vec \rho}_{\rm{o}}}} )\\[4pt]
		
		+\frac{1}{16}{e^{4i\phi}}\delta ({k - 4{{\vec p}_\varphi } \cdot {{\vec \rho}_{\rm{o}}}} ) 
		+\frac{1}{16}{e^{-4i\phi}}\delta ({k + 4{{\vec p}_\varphi } \cdot {{\vec \rho}_{\rm{o}}}} ),\\
	\end{array}
\end{equation}
respectively. Therefore, it is easy to find that the circle of OTF support with $N$-order SSSIM scheme is $(N+\sqrt N)p$.

\begin{table}[!htbp]
	\centering 
	\caption{The proportion in percentage of each PSD in SSSIM scheme at the case of $N=2, 3, 4$.}\label{table:1}
	\vspace{2pt} 
	\begin{tabular}{c|c|c|c|c|c|c|c|c|c} 
		\toprule[2pt]
		\diagbox[dir=NW]{order}{\%}{FSD}   &-4   &-3   &-2    &-1    &0     &1     &2     &3    &4\\
		\hline \rule{0pt}{14pt}
		N=2 &0    &0    &6.25  &25    &37.5  &25    &6.26  &0    &0\\[5pt]\cline{1-10} \hline \rule{0pt}{14pt}
		N=3 &0    &1.56 &9.38  &23.44 &31.25 &23.44 &9.38  &1.56 &0\\[5pt]\cline{1-10} \hline \rule{0pt}{14pt}
		N=4 &0.39 &3.13 &10.94 &21.88 &27.34 &21.88 &10.94 &3.13 &0.39\\[5pt]
		\toprule[2pt]
	\end{tabular}
\end{table}
As we can see, different order $N$ in SSSIM scheme leads to changes in the proportion of various FSDs. Table~\ref{table:1} shows the proportion in percentage of each PSD in the case of $N=2,3,4$. Here, the total percentage is obtained by summing the absolute values of the weights of each PSDs in the Fourier transform of modulation terms. It is worth noting that the proportion of $\text{FSD}_N$ in the $N$-order SSSIM scheme is small, such as the percentage of $\text{FSD}_4$ is only 0.39\%. Therefore, a low sampling and high image quality algorithm is key for the auto-correlation of light intensity.

\section{Conclusion} \label{Conclusion}
In conclusion, we proposed theoretically and demonstrated numerical experimentally the sub-Rayleigh second-order SSSIM scheme. The spatial resolution can surpass the Rayleigh resolution limit by a factor of $2+\sqrt{2}$. The SSSIM scheme is effective whether for 0-1 binary object or natural grayscale object. What's more, the spatial resolution with $N$-order SSSIM scheme can surpass the Rayleigh resolution limit by a factor of $N+\sqrt{N}$. However, the suppression of noise plays a key role to ensure a high signal-to-noise ratio for the $\text{FSD}_N$ signal. 

\section*{acknowledgements}
Liming Li thanks Qiulan Liu for helpful discussions. This work is financially supported by the National Natural Science Foundation of China (NSFC) (62105188).

\end{document}